\documentclass[11pt,twoside]{article}

\usepackage{asp2014}

\aspSuppressVolSlug
\resetcounters

\begin{document}

\title{GPU acceleration of the SAGECal calibration package for the SKA}

\author{Hanno Spreeuw,$^1$ Ben van Werkhoven,$^1$ Sarod Yatawatta,$^2$ and Faruk Diblen$^1$
\affil{$^1$Netherlands eScience Center, Amsterdam, The Netherlands; \email{h.spreeuw@esciencecenter.nl}}
\affil{$^2$ASTRON, The Netherlands Institute for Radio Astronomy, Dwingeloo, The Netherlands}}

\paperauthor{Hanno Spreeuw}{h.spreeuw@esciencecenter.nl}{0000-0002-5057-0322}{Netherlands eScience Center}{}{Amsterdam}{Noord-Holland}{1098 XG}{The Netherlands}
\paperauthor{Ben van Werkhoven}{b.vanwerkhoven@esciencecenter.nl}{0000-0002-7508-3272}{Netherlands eScience Center}{}{Amsterdam}{Noord-Holland}{1098 XG}{The Netherlands}
\paperauthor{Sarod Yattawatta}{yattawatta@astron.nl}{0000-0001-5619-4017}{Astron}{Research and Development}{Dwingeloo}{Drenthe}{7991 PD}{The Netherlands}
\paperauthor{Faruk Diblen}{f.diblen@esciencecenter.nl}{0000-0002-0989-929X}{Netherlands eScience Center}{}{Amsterdam}{Noord-Holland}{1098 XG}{The Netherlands}



\begin{abstract} 

SAGECal has been designed to find the most accurate calibration solutions for low radio frequency imaging observations, with minimum 
artefacts due to incomplete sky models. SAGECAL is developed to handle extremely large datasets, e.g., when the number of frequency bands greatly 
exceeds the number of available nodes on a compute cluster. Accurate calibration solutions are derived at the expense of large computational loads, which 
require distributed computing and modern compute devices, such as GPUs, to decrease runtimes. In this work, we investigate if the GPU 
version of SAGECal scales well enough to meet the requirements for the Square Kilometre Array and we compare its performance with the CPU version.

\end{abstract}

\section{Introduction}
The twenty-first century has given birth to a series of sensitive radio telescopes with designs radically different from those of the previous century. The continuous ambition to look still further back in time and detect the faintest signals at low frequencies has forced antenna engineers to step away from dishes as the collectors of these signals. The reason for this is that steerable dishes would have to become impractically large, at least for long wavelengths \citep{2005ITAP...53.2480E}. Beamforming arrays with many low-gain elements built in the twentieth century suffered from a number of problems which have only recently been overcome \citep{2005ITAP...53.2480E}.

Maximum sensitivity at low frequencies comes only with large bandwidths, which means increased data rates. Hence, a regular observation from one of these modern instruments has a size which cannot be reduced easily on a single computer. Converting the uncalibrated visibilities from an observation into a sky image requires a number of compute nodes on a cluster for a number of days \citep{2017A&A...598A.104S}. This is definitely a paradigm shift with respect to the early 2000's when reducing observations from classical radio telescopes could be done on desktop computers.

The Epoch of Reionization (EoR) Key Science Project of LOFAR \citep{2011AAS...21710704B} tries to detect a very faint 21 cm signal from the early universe, at the time when the first stars and galaxies were formed. This signal is redshifted to frequencies at which LOFAR can observe. This is an example of a LOFAR project that produces massive amounts of data: tens of PetaBytes are foreseen. SAGECal \citep{Yatawatta2012GPUAN}, the software package used to calibrate these observations, was adapted to handle observations that cannot be stored on regular disks in consumer type desktop computers \citep{2018MNRAS.475..708Y}. SAGECAL is a distributed application with CPU, GPU, MPI and Spark support, but can also be deployed as using Docker or Singularity\footnote{\url{https://github.com/nlesc-dirac/sagecal}}.

In this work, we will demonstrate how SAGECal can process increasingly large datasets and observations with a large field of view with respect to the size of the isoplanatic patch of the ionosphere, which affects the required number of directions for calibration. In other words, this work is an investigation of the scalability of SAGECal up to and including datasets that we can expect from the Square Kilometre Array (SKA\footnote{\label{SKAfacts}\url{http://www.skatelescope.org/wp-content/uploads/2018/08/16231-factsheet-telescopes-v71.pdf}}). 

\section{3D comparison between the performance of SAGECal on a CPU and a GPU}

To assess the performance of SAGECal in a quantitative manner, we first constructed five artificial datasets, with 64, 128, 256, 384 and 512 stations. 64 stations corresponds roughly to the number of LOFAR stations (51 presently\footnote{\url{https://www.astron.nl/lofarscience2019/Documents/Monday/LUM_Pizzo.pdf}}, 2 more international stations are coming) while 512 is equal to the planned number of stations for SKA1 LOW\textsuperscript{\ref{SKAfacts}}. Fig.\ref{CPU_GPU_comparison_3D} shows the runtimes for calibration of these five datasets, for SAGECal on a CPU as well as a on GPU. We also included a third axis, the number of clusters: for any input sky model one can group contiguous sources into clusters and a single calibration solution will be derived for each cluster. It is clear that the GPU accelerates SAGECal calibration for all but the lowest workload: 64 stations and just one cluster. In that case, the GPU is redundant and mostly idle; data transfers to and from the GPU cause a delay compared to the CPU run. For all other cases, short GPU compute times compensate the PCI bus data transfer times, since the GPU can offer more FLOPS than the CPU, hence a higher throughput of the data.

\articlefigure[width=1.0\textwidth]{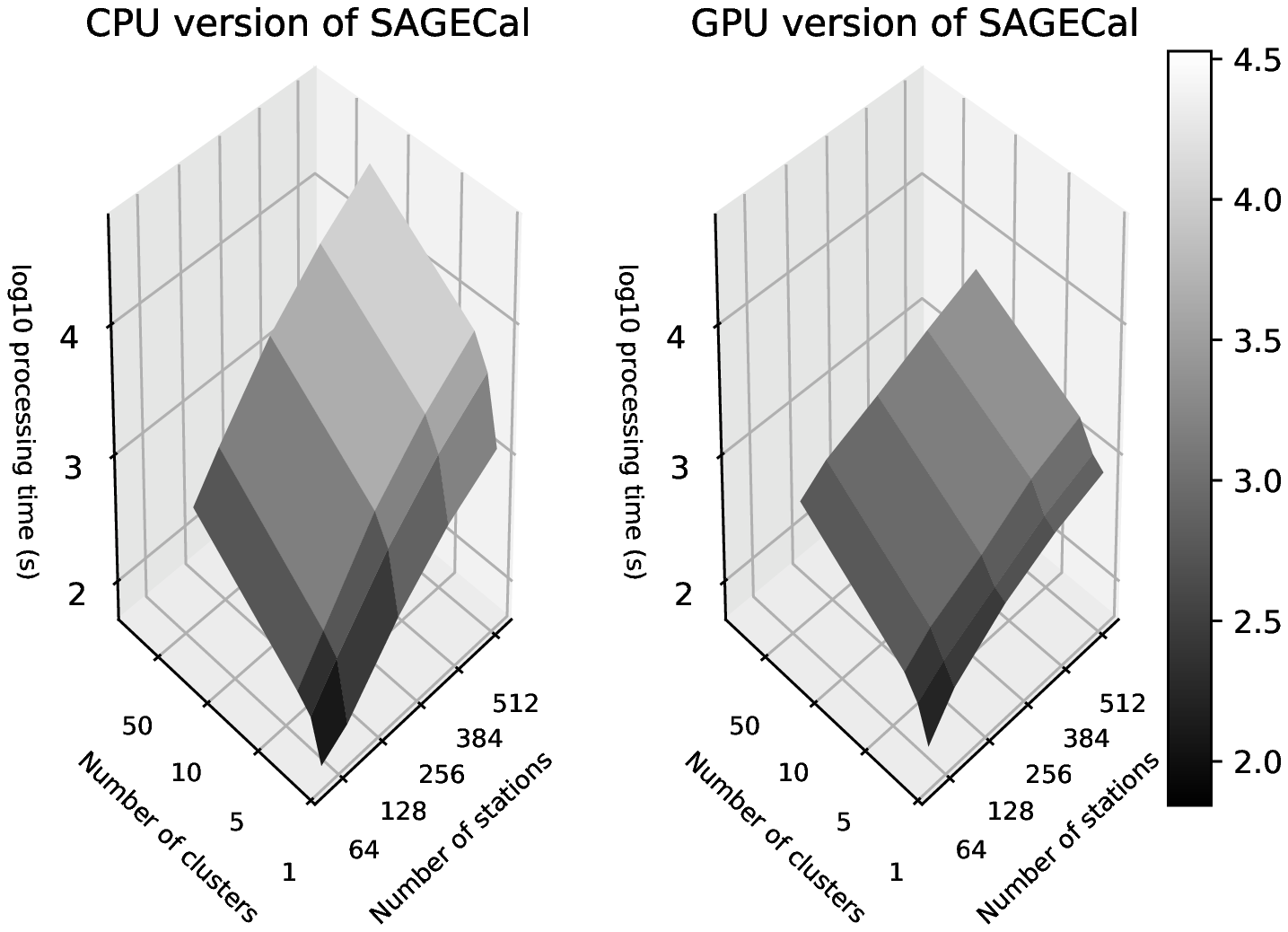}{CPU_GPU_comparison_3D}{We measured the times SAGECal needs to calibrate five artificial data sets, having 64, 128, 256, 384 and 512 stations and for five different sky models, with varying numbers of clusters (calibration directions). We have used a cluster node equipped with two CPUs (2x Xeon E5-2660v3/2640v3, 40/32 logical cores, but number of CPU threads = 32 in all runs) and a NVIDIA GeForce GTX 1080 GPU on ASTRON's DAS5 (\url{https://www.cs.vu.nl/das5/ASTRON.shtml}). }

\section{SAGECal performance on a CPU and on a GPU; a few selected 2D plots}
To get a clearer view of the difference in performance between SAGECal on a CPU and on a GPU, we highlight a few slices from Fig.\ref{CPU_GPU_comparison_3D}. First, we fix the number of calibration directions at 50. The result is given in the left panel of Fig.\ref{CPU_GPU_comparison_2D}. Here, we see a strong non-linear dependence of the CPU runtimes on the number of stations. The size of an observation scales with the number of baselines - $N*(N-1)/2$, with $N$ the number of stations - i.e., a quadratic dependence. While the CPU runtimes seem to scale with data size, the GPU runtimes show a close to linear dependence on $N$. 

While the left panel of Fig.\ref{CPU_GPU_comparison_2D} is clear about a non-linear dependence of the CPU version of SAGECal on the number of stations, it is worthwhile to zoom in on the GPU version to detect any quadratic dependence. The right panel of Fig.\ref{CPU_GPU_comparison_2D} indeed shows that such a dependence, albeit small, may be present, for the datasets we used. Hence, we will need to profile SAGECal on larger datasets to reveal where our algorithms can be improved.

Finally, in Fig.\ref{SAGECal_GPU_runtime_number_of_directions_dependence}, we display the dependence of the runtimes of the GPU version of SAGECal on the number of clusters in the applied sky model, i.e., the number of calibration directions, by fixing the number of stations to 512. It appears that the sampling of calibration directions has been rather coarsely. However, we see no compelling evidence for a strong non-linear component in the runtimes.

\articlefiguretwo{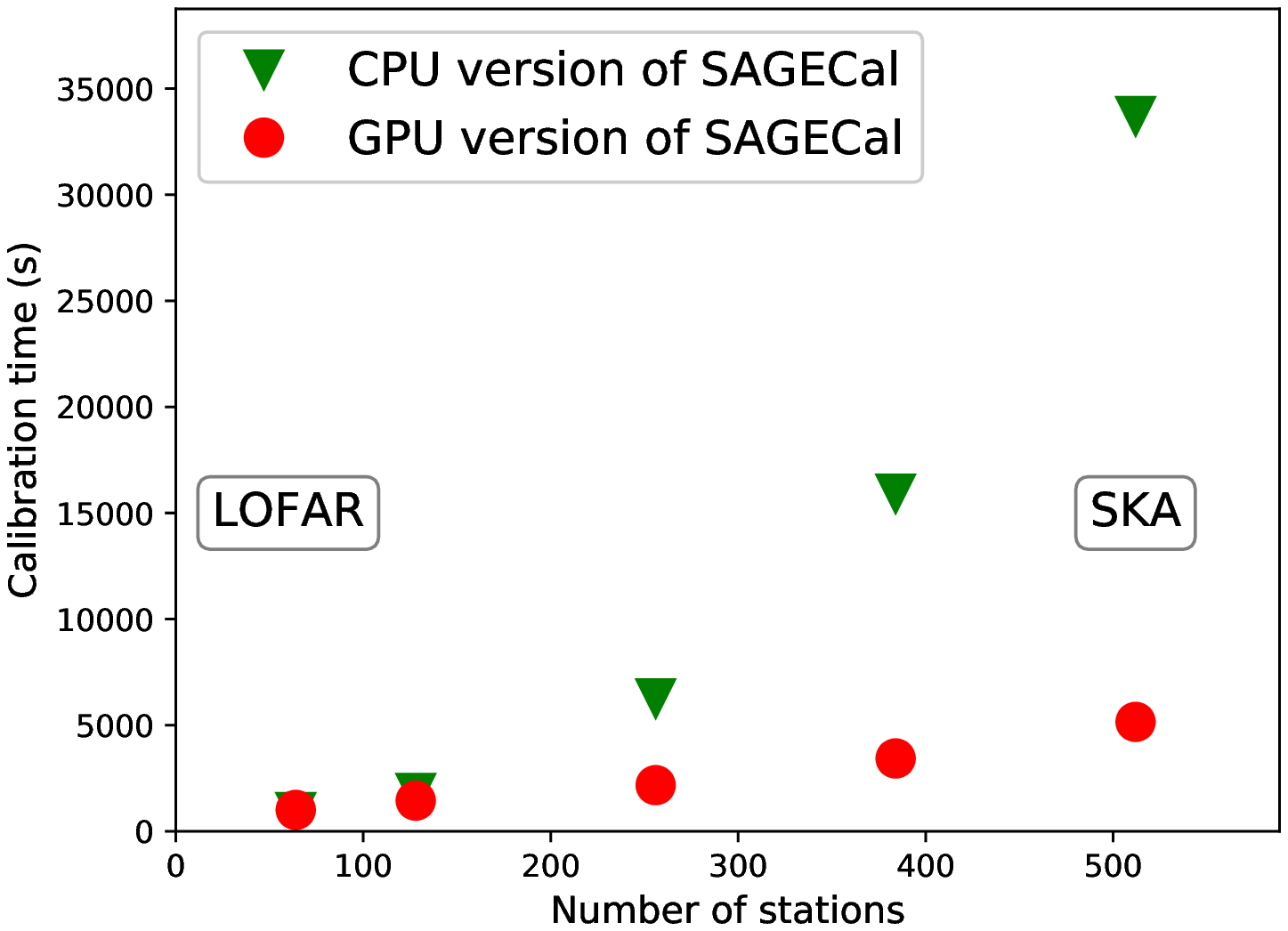}{SAGECal_calibration_times_2D_50_clusters_GPU_only}{CPU_GPU_comparison_2D}{Calibration times for the CPU and GPU versions of SAGECal, for 50 clusters and five numbers of stations. \emph{Left:}  The non-linear dependence on the number of stations is clearly prominent for the CPU version of SAGECal. \emph{Right:} Calibration times from SAGECal on a GPU from the left panel depicted separately. This shows a close to linear dependence on the number of stations, but a small quadratic term may well be present. }

\articlefigure[width=.5\textwidth]{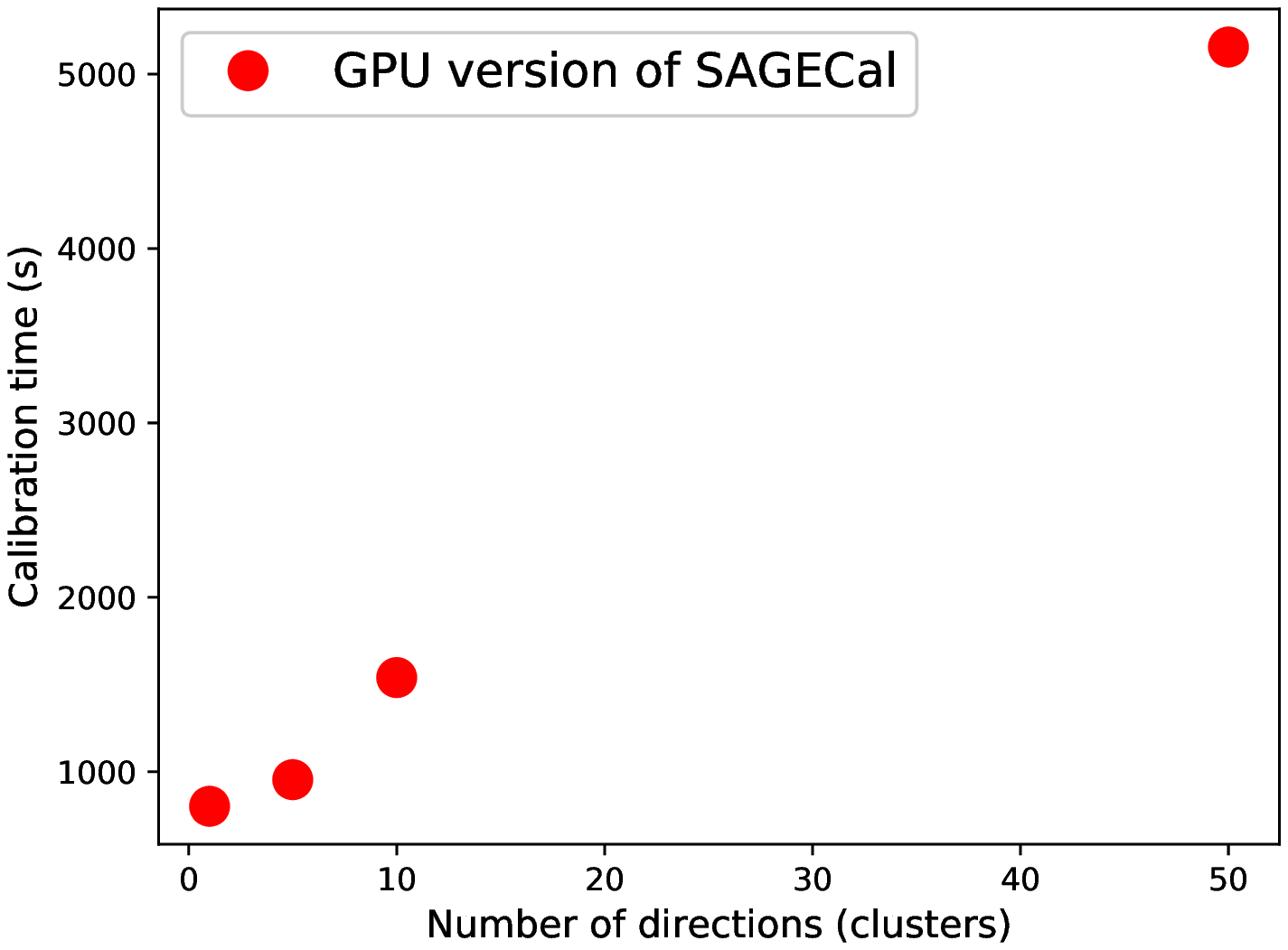}{SAGECal_GPU_runtime_number_of_directions_dependence}{SAGECal on a GPU: dependence of runtime on number of calibration directions with 512 stations.}

\section{Conclusions}  We have investigated to what extent the SAGECal calibration package for interferometric imaging observations at low radio frequencies is "SKA ready", by running a number of tests, both on a CPU and a GPU, on artificial datasets from five different numbers of stations. The dataset from the lowest number of stations (64) corresponds roughly to the present setup of LOFAR while the dataset with the largest number of stations (512) could come from a future observation with SKA1 LOW. We also recorded the SAGECal runtime dependence for four different numbers of calibration directions. Contrary to the CPU version, we find that SAGECal calibration on a GPU scales almost linearly with the number of stations and the number of directions within the ranges expected for SKA. We also found that SAGECal runs about 35\% faster on a low power GPU (NVIDIA Jetson Nano) than on a high end CPU with 40 logical cores (dual Intel Xeon E5-2660v3@2.60GHz) while consuming about twenty times less power.

\acknowledgements This work is supported by The Netherlands eScience Center (project DIRAC, grant 27016G05).

\bibliography{refs}  

\end{document}